# X-ray coherent diffraction imaging with an objective lens: towards 3D mapping of thick polycrystals

Anders Filsøe Pedersen[1], Virginie Chamard[2], Carsten Detlefs[3], Tao Zhou[3], Dina Carbone[4], Henning Friis Poulsen[1*]

1: Department of Physics, Technical University of Denmark, 2800 Kongens Lyngby, Denmark
2: Aix Marseille Univ, CNRS, Centrale Marseille, Institut Fresnel, Marseille, France
3: European Synchrotron Radiation Facility, 71 Avenue des Martyrs, 38000 Grenoble, France
4: MAX IV Laboratory, Fotongatan 2, 225 92 Lund, Sweden
*: corresponding author, hfpo@fysik.dtu.dk

## Abstract

We report on a new x-ray imaging method, which combines the high spatial resolution of coherent diffraction imaging with the ability of dark field microscopy to map grains within thick polycrystalline specimens. An x-ray objective serves to isolate a grain and avoid overlap of diffraction spots. Iterative oversampling routines are used to reconstruct the shape and strain field within the grain from the far field intensity pattern. The limitation on resolution caused by the finite numerical aperture of the objective is overcome by the Fourier synthesis of several diffraction patterns. We demonstrate the method by an experimental study of a ~500 nm Pt grain for the two cases of a real and a virtual image plane. In the latter case the spatial resolution is 13 nm rms. Our results confirm that no information on the pupil function of the lens is required and that lens aberrations are not critical.

## Main
With the advent of ever more brilliant synchrotron sources, x-ray coherent diffraction imaging (CDI) has become a powerful tool for 3D mapping of small samples [1-4]. The method retrieves the sample scattering function from a set of coherent x-ray intensity measurements, using computational inversion approaches to determine the phase of the scattered amplitude, which is not directly measured by the detector. The sample image is obtained by back-propagation of the scattered field. For crystalline specimens, the intensity distribution can be measured in the vicinity of a chosen Bragg peak [5-7], resulting in maps of the material density and of the crystalline displacements projected along the Bragg vector. Existing (Bragg) CDI methods do not employ any optical elements between the sample and the detector. Compared to full-field microscopy methods using an objective lens between the sample and the detector, CDI thus avoids limitations due to aberrations and finite aperture of the lens, and spatial resolutions in the 10 nm range can be obtained with strain sensitivity in the order of a few times $10^{-4}$ [8].

The current use of Bragg CDI is, however, limited to (sub-)micrometer sized isolated crystals exhibiting rather homogeneous strain fields. Bragg ptychography extends the method to laterally extended crystals [9]. For studies of thick polycrystalline samples, none of the above approaches are applicable, because such samples typically contain a large number of simultaneously diffracting micro-crystals within the illuminated volume. Their individual diffraction patterns are incoherently superposed, which makes the phase reconstruction impossible with current methods. Instead 3D grain mapping of mm-sized samples can be carried out by x-ray diffraction tomography methods [10-13] with a resolution of 2 µm. These may be complemented by dark field x-ray microscopy [14-16] where an x-ray objective lens is placed in the Bragg diffracted beam of a selected grain. This magnifies the projection image and acts as a very efficient filter, removing overlap of diffraction from other grains. Lens aberrations, however, currently limit the resolution of dark field microscopy to approximately 100 nm.

Here, we experimentally demonstrate a method that is capable of combining the spatial resolution of Bragg CDI with the ability of dark field x-ray microscopy to study a selected, deeply embedded grain or domain within a thick polycrystalline sample. The method, Objective based Bragg CDI, OBCDI, can be combined with the coarser grain mapping methods described above, enabling multiscale microscopy in 3D for hard crystalline materials. As illustrated in Fig. 1, we use the dark field microscope to create a real or virtual image of the diffracting object. The diffraction pattern of this image is then recorded in the far field Fraunhofer limit. As numerically shown in [17] on a phantom, the phase field introduced by the lens does not modify the intensity distribution in the far-field of the sample (except scaling). This allows the use of usual iterative Bragg CDI algorithms to retrieve the sample crystalline properties [17]. In this experiment, we further show that the spatial resolution limitation introduced by the finite numerical aperture of the objective can be overcome by the introduction of Fourier synthesis approaches [18], applied for the first time in the x-ray regime.

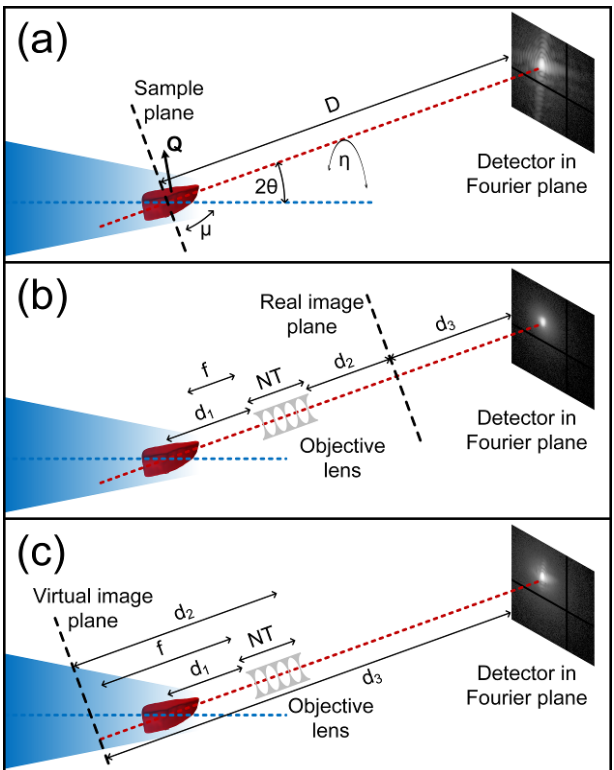

**Figure 1**. Optics principle. (a) Classical Bragg CDI: The detector monitors the far-field diffraction produced by the sample, with no intervening optical elements. The optical axis of the diffracted beam is characterized by ($2\theta$, $\eta$). (b) Objective based CDI with a real image plane (working distance $d_1$ greater than focal length $f$). The objective is a compound refractive lens with N lenslets. (c) Objective based CDI with a virtual image plane ($d_1 < f$). In all cases, the 3D reconstructions are based on images acquired at a set of tilt angles, µ.

The demonstration experiment was carried out at beamline ID01 of the European Synchrotron Radiation Facility, ESRF, using a compound refractive lens (CRL) [19] as objective. Selecting a Pt grain with size ~500 nm within a thin Pt foil, rather than buried in a bulk polycrystal, allows a direct comparison of our OBCDI method with classical Bragg CDI by translating the objective in and out of the diffracted beam, respectively. The aim of this demonstration experiment is to detect and characterize any deviations arising from introducing the objective. In all three configurations, the full 3D intensity pattern is derived from a set of 2D patterns acquired during a "rocking-scan", where the sample is tilted at equidistant angular positions around the axis µ, perpendicular to the incoming beam and the Bragg vector (see Fig. 1).

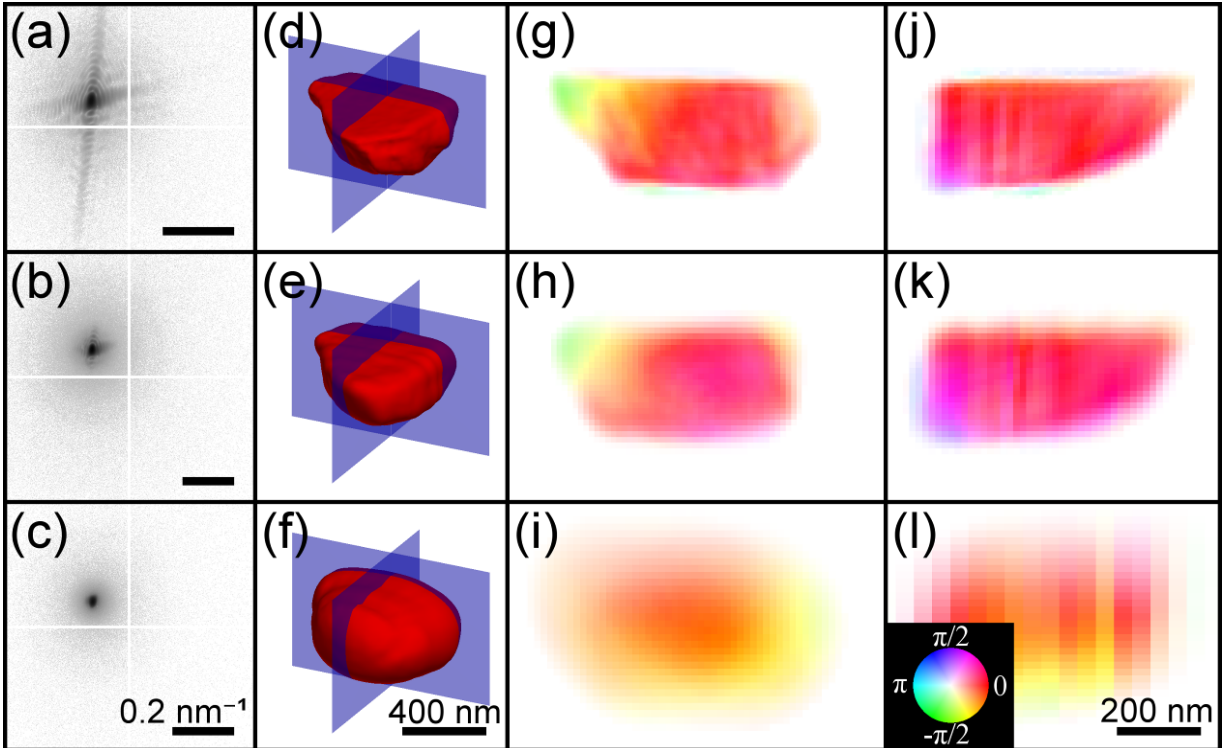

**Figure 2**. Reconstruction of a single Pt grain obtained without Fourier synthesis. Raw data taken at the maximum of the rocking curve (first column), resulting 3D shapes of the selected grain (second column) and phase reconstructions for the two blue planes indicated (third and fourth column). The rows represent classical CDI (top), objective CDI with a virtual image plane (middle) and objective CDI with a real image plane (bottom). The scale bars in (a-c) have the same Fourier space length indicated in (c), a common scale bar for (d-f) is shown in (f) and a common scale bar for (g-l) in (l). The images show the reconstructed complex fields, with the phase represented by the color and the amplitude indicated by the intensity (hue), see the color wheel inset in (l).

Corresponding raw data for the three configurations are shown in Fig. 2. They are similar in the three configurations, with the exception of the anticipated cut-offs caused by the different numerical apertures, cf. Table 1. The resulting reconstructions demonstrate good agreement between the no-objective case and OBCDI. The spatial resolution (rms) is estimated from an error function fit of the surface and is 6 nm, 18 nm and 60 nm, respectively, in the plane shown in Fig. 2g-i. In the Supplementary Information (SI), we show that the differences can be explained almost solely by the band pass defined by the aperture (Fig. S1). This is corroborated by the fact that there is no noticeable difference along the third direction in reciprocal space, the rocking direction, where there is no cut-off. Obtaining 18 nm resolution clearly shows that this method is not sensitive to the CRL's aberrations. This experimentally confirms the results in [17].

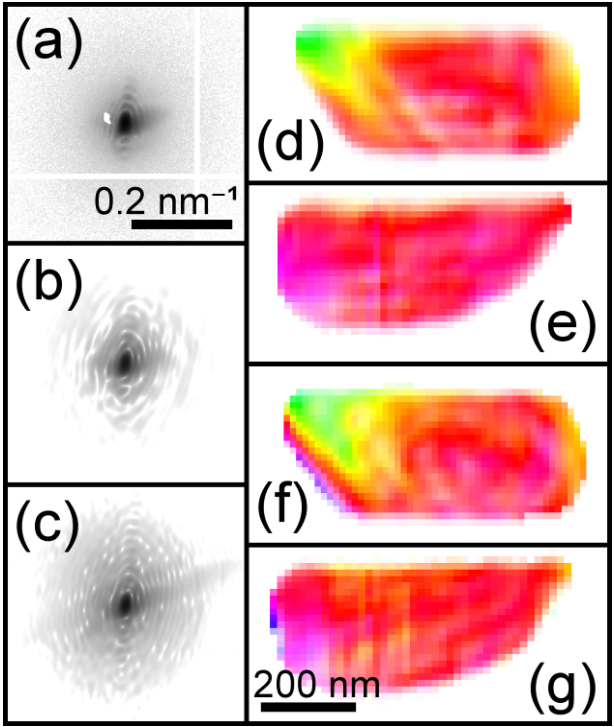

**Figure 3.** Reconstructions obtained using synthesis in Fourier space for OBCDI with a virtual image plane. (a)-(c) Images at the center of the rocking curve ($\mu = 0°$), (a) raw data from a single dataset, (b) the reconstructed Fourier intensity using parallel synthesis, (c) the reconstructed Fourier intensity using serial synthesis. The scale is logarithmic and covers 6 orders of magnitude. (d-e) Reconstructions of two sections in the grain using parallel synthesis, (f-g) reconstructions using serial synthesis, for direct comparison with results in Fig. 2. (a-c) has the same scale, as do (d-g).

To increase the effective bandpass for the virtual image configuration (Fig. 1c), we repeated the data acquisition over a 3x3 grid in the $2\theta$ and $\eta$ directions (see Fig. 1 and Fig. S2 in SI). The grid step size of 0.1° is 0.3 times the numerical aperture of the objective lens, corresponding to a 40-70 % overlap of regions.

Two novel Fourier synthesis algorithms are introduced for the inversion of the nine 3D data sets. In the first one, the data sets are treated in parallel. As an initial seed for all nine independent reconstructions we use the object that was recovered using only the central position of the lens. After convergence of the nine data sets, they are stitched in Fourier space by taking the mean of the aligned reconstructions. The image registration is done using least squares optimization to match the fringes in the intensity measurements, and the shifting is done by the Fourier shift theorem, in order to perform the registration with sub-pixel accuracy. Alignment also involves matching the global phase of the individual data sets, as they are phased separately. This reconstruction method is robust, as it does not rely on any *a priori* information about the pupil size or the exact position of objective. The final result exhibits an rms resolution of 13 nm, as detailed in the SI. More importantly, as seen by comparing Figs. 3d-e with Figs. 2g and 2j, the resulting phase reconstruction becomes remarkably similar to that of classical Bragg CDI.

The second scheme is based on updating sections of the global reconstructed Fourier plane with the measurements in a serial fashion. This method requires knowledge of the pupil function, and the image registration must be done before the reconstruction. The results are shown in Figs. 3f-g, for direct comparison with Figs. 3d-e, 2g, and 2j. Here the resolution is improved to 11 nm (see SI), but artifacts due to experimental instability are showing up, e.g. on the left edge in Fig. 3f. The serial synthesis has more difficulty converging, and is not stable enough to allow support updating using the shrinkwrap algorithm, which we attribute to higher sensitivity to inconsistencies between the nine data sets. In this case, we consider the parallel synthesis to provide the better reconstruction, as the experimental drifts introduce too many artifacts in the serial synthesis reconstruction. For both Fourier synthesis methods, we observe that overlap in Fourier space is not necessary for a successful reconstruction.

In conclusion, we have demonstrated a new technique for 3D measurement of shape and strain distribution of individual grains deeply embedded in a polycrystalline material. Compared to state-of-the-art dark field x-ray microscopy, the spatial resolution can be improved by a factor of at least 5 and is only limited by data acquisition time and experimental stability. Moreover, the set-up presented is easily combined with both dark field x-ray microscopy and coarse grain mapping of the entire specimen. The resulting ability to perform multiscale studies is vital for investigating the dynamics of the microstructure in the vast class of structural materials, and as such critical for the ambition of developing materials models of sufficient quality to enable 'materials design in the computer'. This work is a proof-of-concept; the optimization of experimental geometry and reconstruction algorithms as well as first applications on thick specimens will be detailed in future publications. In particular, the recent manufacture of high quality Multilayer Laue lenses [20] makes it possible to employ hard x-ray objectives with a higher numerical aperture [21].

The approach presented shares its roots with e.g. Fourier Ptychography [22] and phase contrast confocal microscopy [23]. It couples direct and Fourier space in several ways, with potential for new applications, in microstructure imaging – which by definition is six-dimensional – and elsewhere. The configuration with the real image plane enables images to be acquired in both direct space and Fourier space (in the back focal plane or in the far field diffraction plane). This enables combined reconstructions and the use of *a priori* knowledge about the sample, such as support constraints, in both domains. We believe such advances could be a route towards higher resolution in both domains and the possibility to extend CDI methods towards samples exhibiting more complex strain variations.

## Methods
The sample is a Pt foil (99.95%, Alfa Aesar) cold rolled to 4 µm thickness. A 3x5 mm$^2$ piece was subsequently annealed at 650°C in air for 5 hours. Partial recrystallization took place, resulting in ~500 nm sized grains with relatively low strain – ideal for Bragg CDI.

The experiment was carried out at beamline ID01 at ESRF at a photon energy of 8 keV. By a combination of a Si(111) Bragg-Bragg monochromator, a slit and a KB mirror used as condenser, a fully coherent incoming beam was generated with dimensions of 1.5 x 2.0 µm$^2$ at the sample position. The sample was mounted on a high precision Huber 3+2 circle diffractometer with a hexapod and a piezo-stage for coarse and fine sample movements, respectively. The removable CRL serving as objective comprised N identical Be biparaboloid lenslets with an apex radius of curvature of 50 µm placed a distance T = 2 mm apart. N was 70 and 20 in the real and virtual image configuration, respectively. The microscope was aligned with the optical axis corresponding to the Pt(111) Bragg peak, at 2θ ~ 40°. The far field detector was a MaxiPix detector with 516x516 pixels of 55 µm size, mounted at the end of a detector arm. An evacuated flight tube was placed between the CRL (or sample in the case of no objective) and the detector to minimize air absorption and scattering.

In all cases we used a rocking step size of Δµ = 0.003°. For the no objective and the virtual image plane configurations we used 281 steps, for the real image plane configuration 101 steps. Key geometrical parameters are summarized in Table 1. In all experiments, the sample-detector distance was fixed at 1.632 m. Using geometrical optics expressions for a thick lens [15] the distances were configured to match a magnification of 1.4 and -0.9, respectively. The actual magnification was determined with high precision by comparing to the fringe pattern to the Bragg CDI case.

|  | N | $d_1$ | $d_2$ | $d_3$ | M | NA | $D_{eff}$ | $\sigma_{pupil}$ | $N_F$ |
|---|---|---|---|---|---|---|---|---|---|
| Bragg CDI | - | - | - | - | - | ~13.5 | 1.632 | - | - |
| OBCDI, real image | 70 | 0.121 | 0.100 | 1.271 | -0.899 | 1.44 | 1.414 | 7.85 | 0.013 |
| OBCDI, virtual image | 20 | 0.050 | -0.119 | 1.661 | 1.409 | 5.5 | 1.179 | 25.2 | 0.032 |

**Table 1**. Key geometrical parameters. Distances are defined in Fig. 1 and are in meters. The numerical aperture (NA) represents the FWHM of the detectors' (for Bragg CDI) or CRL's (for OBCDI) angular acceptance and is in mrad. M is the magnification. $D_{eff} = d_3/M$ is an effective sample-detector distance, the distance at which the intensity patterns on a hypothetical detector have the same scale as the Bragg CDI configuration. $\sigma_{pupil}$ gives the rms width of the Gaussian pupil function in units of pixels. $N_F$ is the Fresnel number at the entrance of the lens, assuming a 500 nm aperture (sample). This shows that the objective lens is placed in the far field limit.

The reconstructions without synthesis were performed by using a combination of the hybrid input-output algorithm (HIO) [24] with a feedback parameter of β = 0.9, the error reduction algorithm (ER) [25], and the shrinkwrap algorithm [26] to update the support. The initial guess of the object was the inverse Fourier transform of the measured Fourier amplitudes with a random phase, and the initial support was calculated from the autocorrelation function. 100 cycles of 20 x HIO + 1 x ER + shrinkwrap were repeated, followed by 100 cycles of 20 x HIO + 1 x ER, where the object was average after every cycle following [27].

For the parallel Fourier synthesis, we reconstructed the phase in the Fourier plane independently for the eight additional datasets, and then synthesized the complex fields in the Fourier plane. First, the image registration of the intensity patterns was determined by fitting the rocking curve and employing a mean square difference optimization. The shifts within the images were determined using the Fourier shift theorem, and were therefore determined to a sub-pixel resolution. The intensity patterns were also normalized to match the Fourier amplitude between each dataset. As an initial guess, the object reconstructed from the single data set at the center of the grid was used, as well as the support. The ER algorithm was used to reconstruct the phase in the Fourier plane, and after each step a combined object was formed by averaging the objects. The shrinkwrap algorithm was used on this combined object. After 100 cycles the reconstruction had converged, and the Fourier reconstructions were aligned and averaged. Alignment also implied matching the global phase in the nine data sets, as these were phased individually, while the phase retrieval algorithm only determines the relative phase within a data set. The final object was found by inverting this synthesized Fourier amplitude and phase. In this experiment we had slight sample drifts, impacting the position of the pupil function for each dataset, making the alignment of the data set critically important.

The serial Fourier synthesis was carried out by updating the amplitude in Fourier space, weighted by the Gaussian pupil function, with the measured intensities for one data set at a time. The ER algorithm was used to enforce the support in between each updating cycle. Again, the initial seed was the reconstruction from the central data set. At the end, the object was directly recovered, in contrast to the parallel synthesis method.

Interestingly, we noticed that the convergence was still ensured if only a data subset extracted from the nine data sets was used (e.g., four instead of nine). As one expect, it resulted however in a slightly degraded solution due to the lower number of photons in the subsets. Those tests demonstrated that the overlap in the Fourier domain does not bring additional information for the phase retrieval. The inversion was made possible by the existence of a finite support in the real space, making the link between our approach and CDI.

**Acknowledgements**
We acknowledge funding support from ERC Advanced Grant nr. 291321. We also acknowledge funding from DanMAX, grant number 4059-00009B. We thank Yubin Zhang and Jacob Bowen for help with providing the sample, Hugh Simons, Mario Beltran and Jens Wenzel Andreasen for scientific discussions, ESRF for provision of beamtime (MI-1315) and the instrument center Danscatt for a travel grant.

**Author Contributions**
AFP analyzed the data with input from VC. HFP conceived the original idea. AFP, VC and TZ conceived and designed the experiment. TZ, AFP, CD, VC and GC performed the experiment. HFP, AFP, VC and CD wrote the paper.

# Supporting information: X-ray coherent diffraction imaging with an objective lens: towards 3D mapping of thick polycrystals

Anders Filsøe Pedersen[1], Virginie Chamard[2], Carsten Detlefs[3], Tao Zhou[3], Dina Carbone[4], Henning Friis Poulsen[1*]

1: Department of Physics, Technical University of Denmark, 2800 Kongens Lyngby, Denmark
2: Aix Marseille Univ, CNRS, Centrale Marseille, Institut Fresnel, Marseille, France
3: European Synchrotron Radiation Facility, 71 Avenue des Martyrs, 38000 Grenoble, France
4: MAX IV Laboratory, Fotongatan 2, 225 92 Lund, Sweden
*: corresponding author, hfpo@fysik.dtu.dk

## Simulation of the effect of the aperture

We can simulate the effect of the pupil function on the classical (lensless) Bragg CDI data. To this end we Fourier transform the reconstruction for the classical BCDI – shown in Fig. S1a – multiply by the Gaussian pupil function with the width specified in Table 1 in the main text, and then inverse Fourier transform the result back to real space. The results using the pupil functions for the virtual and real image geometry are displayed in Figs. S1b-c, respectively. These results can be compared directly to the reconstructions with the objective in place, shown in Figs. 2h-i, and for convenience reproduced here as Figs. S1d-e. It is evident that the discrepancy between the classical BCDI result and the OBCDI results to a large extent can be explained by the finite pupil.

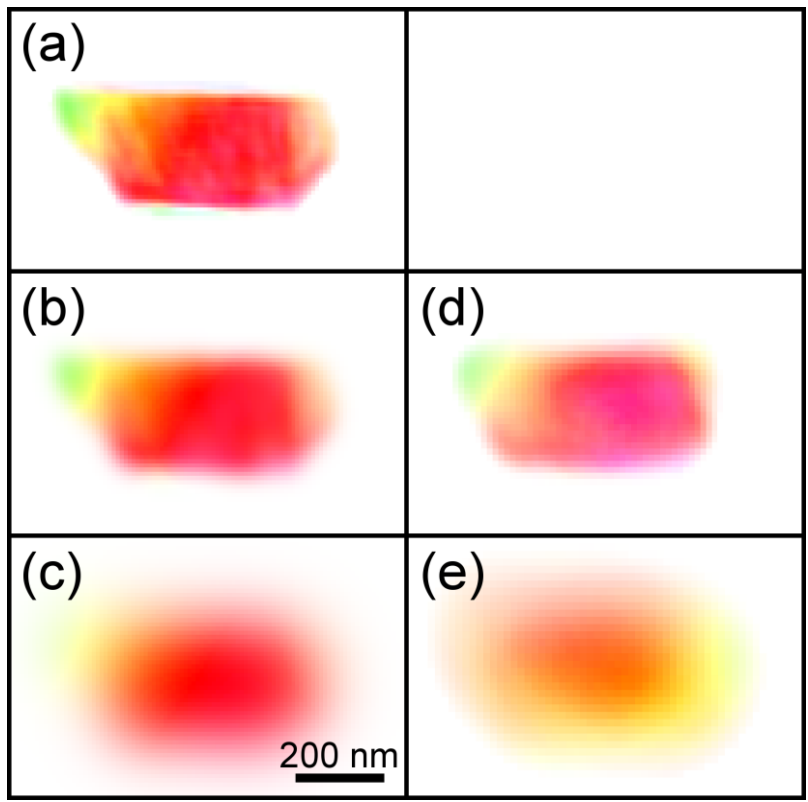

**Figure S1**. A study of the effect of the finite pupil on reconstruction quality. The classical BCDI reconstruction in (a) is used as a phantom for numerical studies. Results from applying a pupil of size corresponding to the virtual image case (b), and the real image case (c) are compared to the reconstructions from the measurements with the lens in the virtual and real image geometry in (d) and (e), respectively. All plots share the same scale.

Measurements for Fourier synthesis

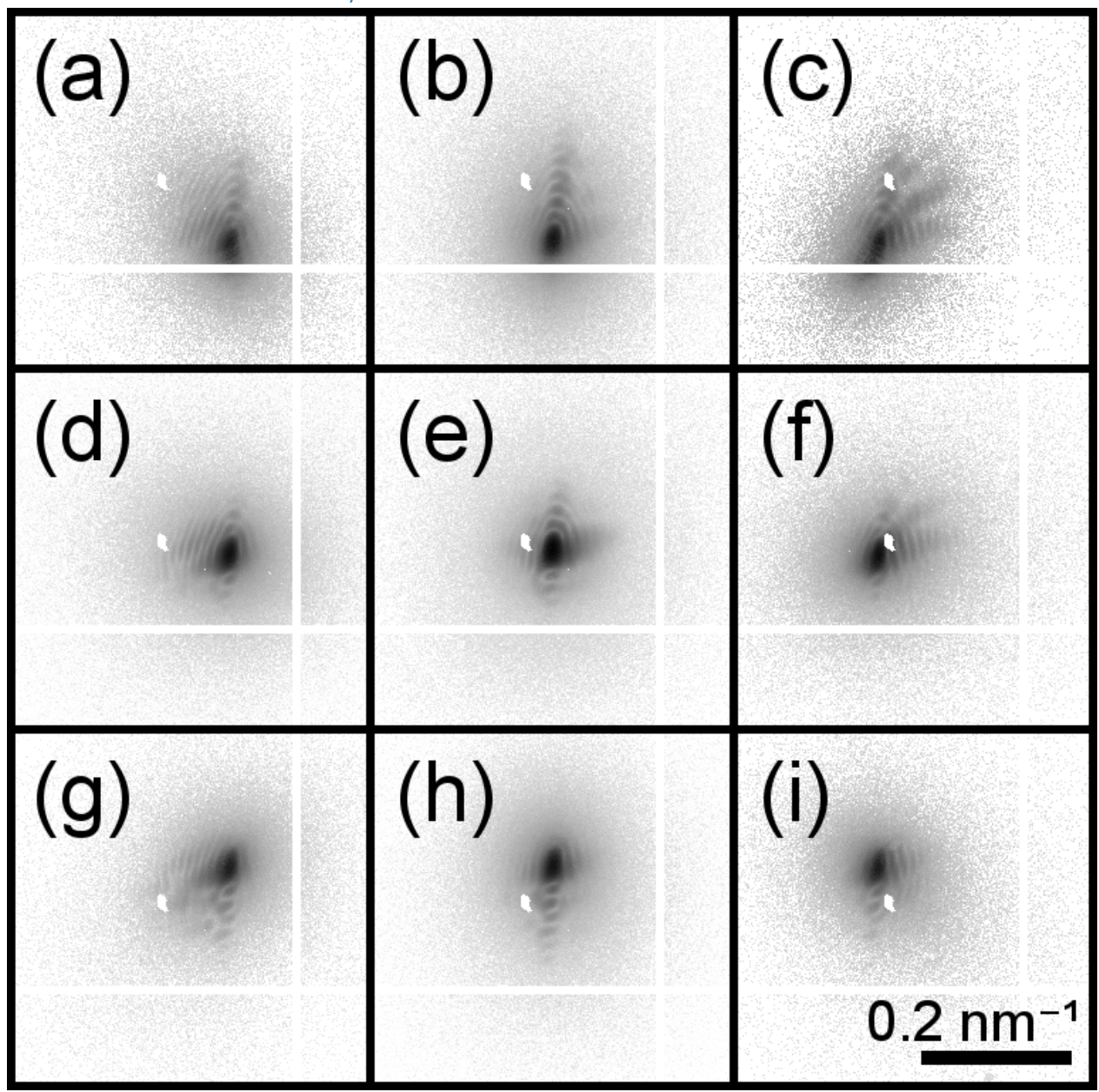

**Figure S2**. The intensity pattern recorded at the central Bragg peak for the nine data sets used for Fourier synthesis. The images are plotted on a logarithmic scale covering six orders of magnitude. The rows correspond to steps in 2θ and the columns to steps in η angle. The scale of all the images is the same with a common scale bar shown in (i).

Shown in Fig. S2 are corresponding examples of raw data for the nine data sets used for the Fourier synthesis in the virtual image geometry. The step size in 2θ and η are both 0.1° steps. The white lines are gaps between the detector modules.

## Fourier synthesis resolution

The resolution of the two Fourier synthesis methods is evaluated by comparing the extent of the intensity in Fourier space. Fig. S3 shows the normalized mean intensity as a function of reciprocal space length in the detector plane. Compared to the reconstruction based on a single data set only, the extent of the Fourier intensity at a relative level of $10^{-3}$ is increased by 36% and 64% for the parallel and serial Fourier synthesis, respectively. This translates into improving the resolution from 18 nm to 18/1.36 = 13 nm and 18/1.64 = 11 nm, respectively.

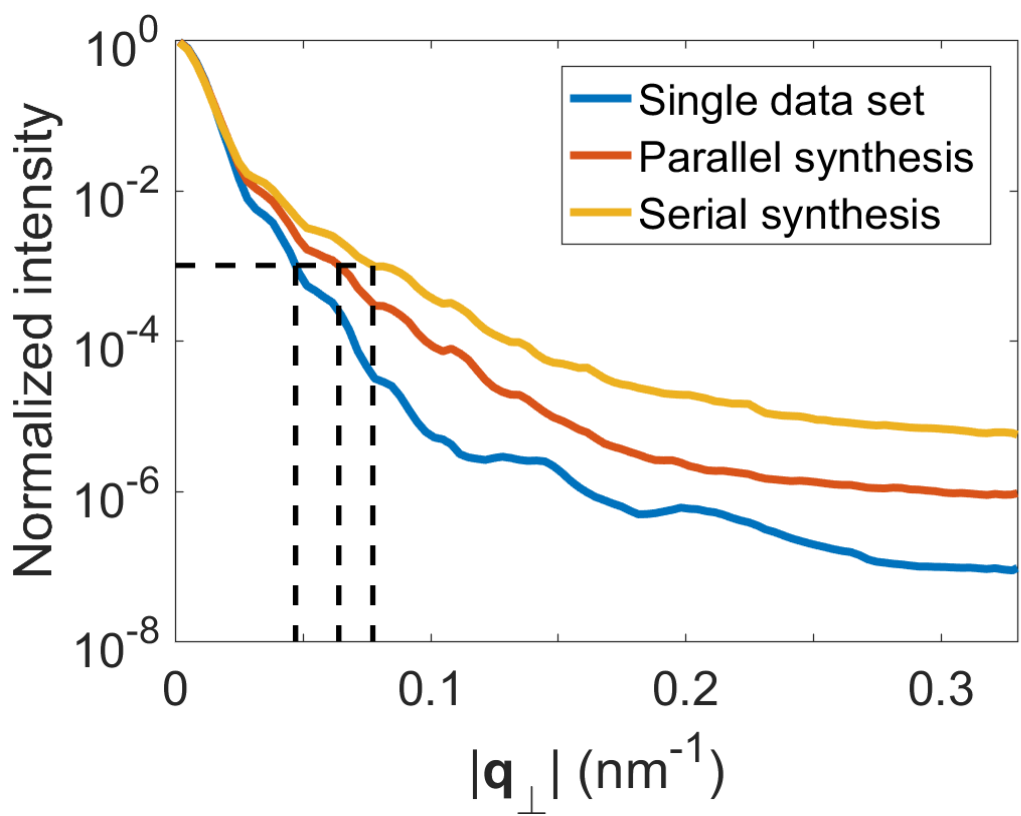

**Figure S3**. The averaged and normalized intensity for the single data set and the two Fourier synthesis methods are plotted against the q-vector magnitude in the detector plane.